\documentclass[aps,prb,preprint,groupedaddress,superscriptaddress, amsmath,amssymb, aps,floatfix, showpacs]{revtex4-1}
\usepackage[final]{graphicx}
\usepackage{hyperref}
\usepackage{ifpdf}
\begin{document}

\title{Magnetic Structure in Fe/Sm-Co Exchange Spring Bilayers with Intermixed Interfaces}

\author{Yaohua~Liu}
\email[]{yhliu@anl.gov}
%\affiliation{Materials Science Division, Argonne National Laboratory, Argonne, IL 60439, USA}
\author{S.~G.~E.~te~Velthuis}
%\affiliation{Materials Science Division, Argonne National Laboratory, Argonne, IL 60439, USA}
\email[]{tevelthuis@anl.gov}
\author{J.~S.~Jiang}
%\affiliation{Materials Science Division, Argonne National Laboratory, Argonne, IL 60439, USA}
\author{Y.~Choi}
\altaffiliation[Current address:~]{Advanced Photon Source, Argonne National Laboratory, Argonne, IL 60439, USA}

\author{S.~D.~Bader}
\affiliation{Materials Science Division, Argonne National Laboratory, Argonne, IL 60439, USA}

\author{A.~A.~Parizzi}
%\affiliation{Spallation Neutron Source Oak Ridge National Laboratory, Oak Ridge, TN 37831, USA}

\author{H.~Ambaye}
\author{V.~Lauter}

\affiliation{Spallation Neutron Source Oak Ridge National Laboratory, Oak Ridge, TN 37831, USA}

\date{\today}

\begin{abstract}
 The depth profile of the intrinsic magnetic properties  in an Fe/Sm-Co bilayer fabricated under nearly optimal spring-magnet conditions was determined by complementary studies of polarized neutron reflectometry and micromagnetic simulations. We found that at the Fe/Sm-Co interface  the magnetic properties change gradually at the length scale of 8~nm.  In this intermixed interfacial region, the saturation magnetization and magnetic anisotropy are lower and the exchange stiffness is higher than values estimated from the model based on a mixture of Fe and Sm-Co phases. Therefore,  the intermixed interface yields superior exchange coupling between the Fe and Sm-Co layers, but at the cost of average magnetization.

\end{abstract}
%75.70.Cn 	Magnetic properties of interfaces (multilayers, superlattices, heterostructures)
%61.05.fj 	Neutron reflectometry
\maketitle

\section{Introduction}

Exchange-coupled, high-magnetization (soft) and high-anisotropy (hard) magnetic phases have potential applications as both ultra-strong permanent magnet~\cite{KnellerHawigIEEE1991} and ultra-high-density recording media.~\cite{VictoraIEEE2005} While the intrinsic properties of the two phases play the most important role, optimization of the interface properties are also important to achieve the best performance. For example, the interface morphology need to be optimized for good  exchange coupling between the soft and hard phases.  Micromagnetic simulations suggested that magnetically graded interfaces, whose magnetic properties are gradually changed over a distance $\sim 10$~nm, is more effective than sharp interfaces, in order to increase the nucleation field in the soft phase and decrease the switching barrier of the hard phase.~\cite{JiangAPL2004, SuessAPL2006} Magnetically graded interfaces can be fabricated from chemically intermixing phases via nanotechnology.~\cite{ChoiAPL2007, GollJAP2008, KirbyPRB2010} Nanoscale spatial resolution is needed to experimentally determine the intrinsic magnetic properties of hard-soft heterostructures.  Interestingly, for Fe/Sm-Co spring magnets under optimized fabrication conditions for the maximum energy product, the Fe layer and the Sm-Co layer have an intermixed interface over a length scale $\sim 5-10$~nm.~\cite{ChoiPRB2007, LiuAPL2008}  Although the intrinsic magnetic properties of individual Fe films and Sm-Co films have been well studied,~\cite{FullertonPRB1998}  there lacks quantitative knowledge about the intrinsic magnetic properties of the intermixed Fe/Sm-Co interface due to its complex composition.~\cite{ChoiPRB2007, LiuAPL2008}

Polarized neutron reflectometry (PNR) is a sensitive tool to study the depth profile of magnetic structures within multilayers with sub-nanometer depth resolution,~\cite{Fitz2005, Chatterji2006} and has been used to study exchange coupling and anisotropy.~\cite{DonovanPRL2002, FitzPRB2006}  For example, in a NiFe/FePt spring magnet, O'Donovan~\textit{et al.} have determined that the twisted magnetic structure of the spring magnet is not confined to the magnetically soft layer, but also penetrates into the hard magnetic phase.~\cite{DonovanPRL2002} Recently, Kirby~\textit{et al.} showed qualitatively that the structural gradation yields a graded anisotropy in Co/Pd multilayers.~\cite{KirbyPRB2010}

In the present work, we report PNR studies on a Fe/Sm-Co bilayer, which was fabricated under nearly optimized spring magnet conditions. We first confirmed that there is a structurally intermixed interface of $\sim 8$-nm wide between the Fe and Sm-Co layers via combined X-ray reflectometry (XRR) and PNR studies. We also determined the depth profile of the saturation magnetization $M_S$. Furthermore, the profiles of the magnetic anisotropy $K$ and the exchange stiffness $A$ were obtained by analyzing the PNR data with the aid of micromagnetic simulations. The magnetic properties in the intermixed region were compared to a model based on a mixture of Fe and Sm-Co phases. The saturation magnetization is slightly lower than the value estimated from the model, suggesting new compounds formed in the intermixed region. The intrinsic anisotropy is also lower than the value from the model. However, the exchange stiffness is higher, so that the interface efficiently couples the Fe and Sm-Co layers.

\section{Experimental Methods}
\label{Methods}

The Fe/Sm-Co thin film was fabricated via \emph{dc} magnetron sputtering onto an MgO (110) single-crystal substrate with a nominal structure of (10~nm Cr)/(10~nm Fe)/(20~nm Sm-Co)/(20~nm Cr)/MgO. The sample size is $10 \times 10$~mm$^2$.  The Cr (211) buffer layer  and the Sm-Co layer were grown at 400~$^\circ$C, and the Fe layer was grown at 100~$^\circ$C.~\cite{ChoiAPL2007} The Sm-Co layer, nominally Sm$_2$Co$_7$ composition, has an uniaxial, in-plane magnetic easy axis.~\cite{BenaissaIEEE1998}  Magnetic hysteresis loops were obtained by means of vibrating sample magnetometer (VSM). XRR studies were performed with a X-ray diffractometer using Cu $K_{\alpha}$ radiation. PNR experiments were conducted at the SNS at Oak Ridge National Laboratory, at the Magnetism Reflectometer.~\cite{LauterPhysicaB2009} This is a time-of-flight (TOF) instrument with a wavelength band 2~-~5~\AA~and the polarization efficiency of $\sim 98\%$. All experiments  were conducted at room temperature.

\begin{figure}[bh]
	\centering
		\includegraphics[width=0.32\textwidth]{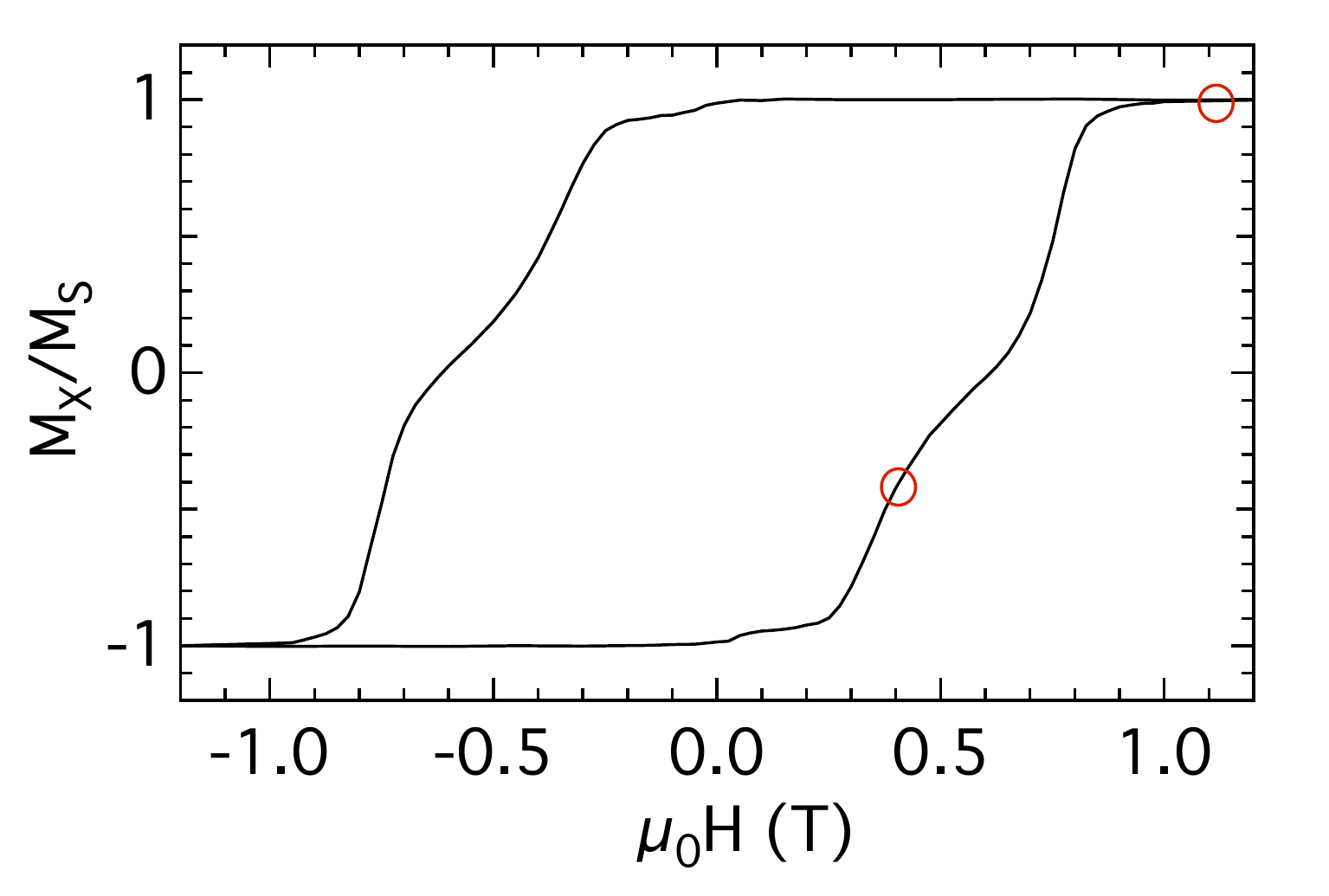}
	\caption{\label{MH}~(Color online) Easy-axis magnetic hysteresis loop. The film is saturated at 1 T. The two circles label the magnetization states measured by PNR, with applied fields of +1.11 and +0.39~T, respectively.}
\end{figure}

The external field was applied along the magnetic easy-axis for PNR experiments. For convenience, we define the field direction  as $+ \hat x $ and the sample surface normal as $-\hat z$.  Figure~\ref{MH} shows the easy-axis magnetic hysteresis loop. The film saturates at fields above $+1$~T.  There are two major reversal processes along the increasing field branch, centered at $+0.35$ and $+0.75$~T, respectively. A minor soft-phase was also observed, which reverses below $+0.05$~T. PNR measurements were performed at $+1.11$~T and in a demagnetized state at $+0.39$~T, after saturation with a $-1.11$~T field, in order to determine the depth profiles of the saturation magnetization, the exchange stiffness and the magnetic anisotropy.

Reflectometry is a non-destructive method to determine scattering length density (SLD) profiles. In the specular condition, XRR yields the depth profile of the electron density, which can be used to reconstruct the chemical structures. Neutrons interact with both nuclei and the internal magnetic field. Spin-up and spin-down neutrons feel the same nuclear scattering potential, but an opposite magnetic scattering potential. From subsequent measurement with oppositely polarized neutron beams, the two contributions can be separated in order to reconstruct both depth profiles of the chemical structure and the magnetization vector.~\cite{Fitz2005, Chatterji2006}  The contrast of the neutron SLDs between Fe and Sm-Co is high, so that the PNR experiments  are sensitive to the interfacial structure of interest.  There are four reflectivities in PNR, including two non-spin-flip (NSF) reflectivities $R^{++}$ and $R^{--}$, and two spin-flip (SF) reflectivities $R^{+-}$ and $R^{-+}$.  SF scattering occurs when the sample's magnetization vector has a non-zero component perpendicular to both the neutron's polarization direction and the momentum transfer direction. During magnetization reversal, the SF scattering occurred in a sufficiently high magnetic field so that it was necessary to take into account the Zeeman effect when analyzing the data.~\cite{FelcherPB1996} (See Appendix~\ref{App:Zeeman}.)

 For PNR, Fredrikze's formalism~\cite{FredrikzePB2001} was used to determine the optics' polarization efficiency and the direct beam (DB) spectra from the measured four reflectivities of the DB. The polarization corrections with the error propagation were made following Wilder's formalism.~\cite{WildesNN2006} Simulations of the reflectivities were based on the Parratt formalism.~\cite{Reflpak} A rough interface was modeled as a sequence of very thin slices whose SLDs vary, following an error function so as to interpolate between adjacent layers.~\cite{Fitz2005} The instrumental resolution was handled by Gaussian convolution. The GenCurvefit program~\cite{GenCurvefit} using the genetic algorithm was employed for the model optimization.

\section{Experimental Results and Analysis}
\label{ExData}

	\subsection{Depth profile of Saturated Magnetization}
	\label{Sec:Corefine}	

\begin{figure}[ht]
	\centering
		\includegraphics[width=0.48\textwidth]{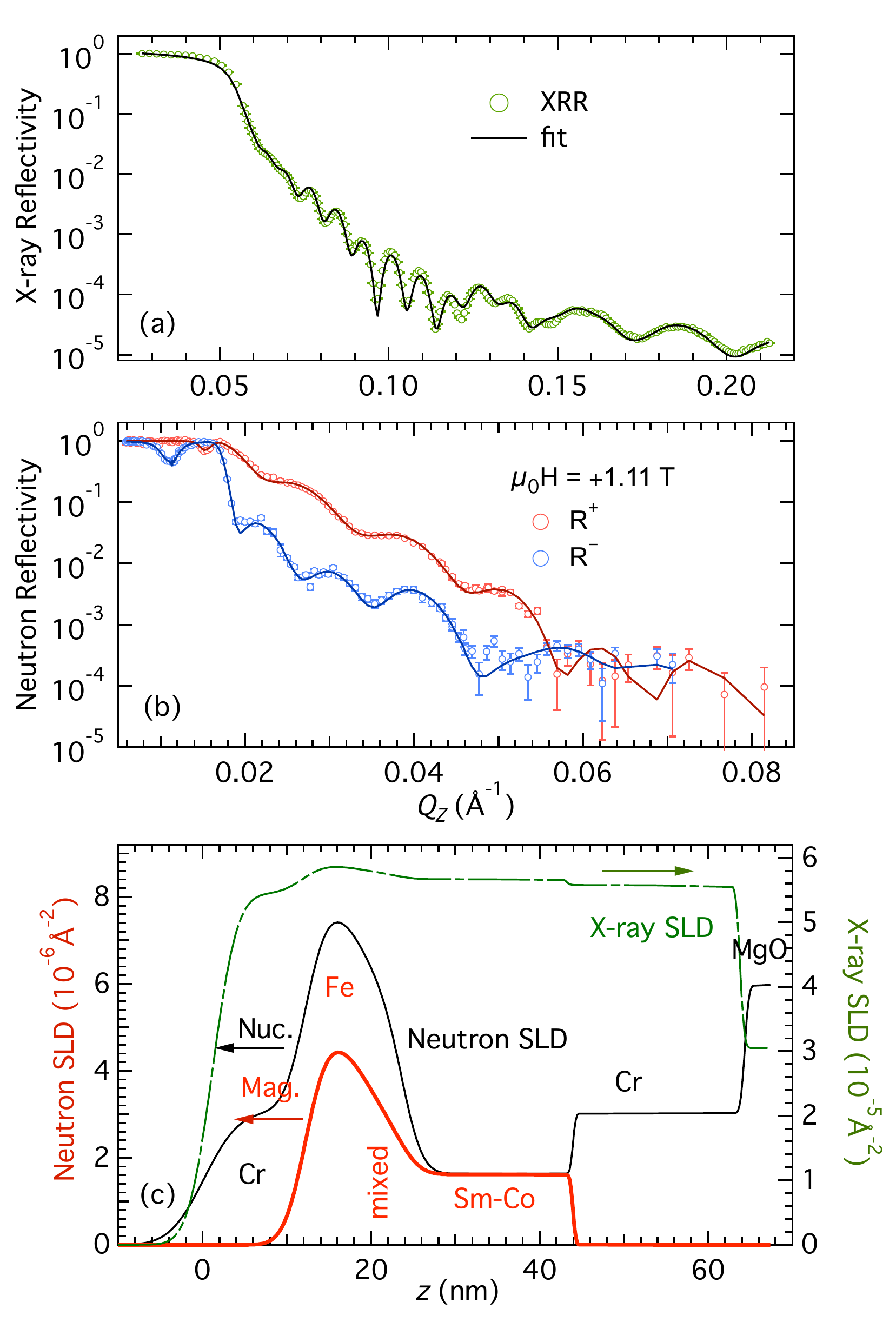}
	\caption{\label{Corefine}~(Color online) (a)~XRR data and (b)~PNR data taken at saturation field $+1.11$~T. (c)~Depth profiles of neutron nuclear SLD (thin black line, left), neutron magnetic SLD (thick red line, left) and X-ray SLD (dashed green line, right). Only the real parts of the X-ray and the nuclear SLDs are shown.}
\end{figure}

The XRR data and the specular PNR data taken at a saturation field of $+1.11$~T are displayed in Figs.~\ref{Corefine}a and ~\ref{Corefine}b, respectively.  The reflectivities are shown as functions of the momentum transfer perpendicular to the film plane $Q_z = 4 \pi sin\theta /\lambda$, where $\theta$ is the incident angle and $\lambda$ is the wavelength of the radiation source. Since there is no spin flip scattering exists at saturation, we measured only $R^{+}$ (=$R^{++}$) and $R^{-}$ (=$R^{--}$). Because of the sample's high saturation magnetization, $R^{+}\gg R^{-}$ for $Q_z$ above the critical edges ($Q_z > 0.018$~\AA$^{-1}$). There is an interesting feature in the PNR data below the critical edges: both $R^{+}$ and $R^{-}$ showed frustrated total reflection. This is due to the enhanced absorption at resonant conditions.~\cite{MaazaPLA1996} Simulations show that the $Q$ separation between the two dips strongly depends on the magnetization along the field direction, especially that  of the Sm-Co layer, which is the locus of absorption. Therefore, this separation, as well as the splitting between $R^{+}$ and $R^{-}$ above the critical edge,  are direct indications of the sample's magnetization.

The chemical and the magnetic SLD profiles were determined by fitting the XRR and the PNR data simultaneously with the same structural model.\cite{Model}  Since there are well tabulated values for the neutron and the X-ray scattering length and absorption length for each element,  both the real and imaginary parts of the X-ray and the neutron's nuclear SLDs were calculated from the chemical composition and the film density. Therefore a single parameter was used to determine both the X-ray and the neutron nuclear SLDs for each layer that has a well defined composition. This parameter is the mass density for that particular layer. However, the intermixed interface has a complex chemical composition that extends $>~5$~nm in depth.~\cite{ChoiPRB2007, LiuAPL2008} The X-ray and the neutron's nuclear SLDs in the intermixed region need to be determined separately, which was done by introducing an intermixed layer between the Fe and Sm-Co layers. Therefore, both the nuclear and the magnetic  SLD profiles can vary in more sophisticated ways and they are less correlated to each other over the intermixed region.  Actually, the peak and dip positions in $R^{-}$ between 0.025 and 0.045 \AA~can not be reproduced without introducing this intermixed layer, which indicates that a single error function is not sufficient to model the nuclear and/or the magnetic SLD profiles at the Fe/Sm-Co interface.

\begin{table*}[hbt]
\caption{\label{tab:SLD} The layer thickness, \emph{rms} interface roughness, the nuclear and magnetic SLDs, saturated magnetization $M_S$, exchange stiffness $A$ and uniaxial anisotropy $K$ in the layers of interest from the best fits. Nuclear and magnetic SLDs and $M_S$ are from fitting the $+1.11$-T data (Sec.~\ref{Sec:Corefine}). $A$ and $K$ are from fitting the $+0.39$-T data (Sec.~\ref{Sec:MM}). Typical literature values are $M_S = 1700~(550)$~emu/cc, $A = 2.8~(1.2)$~ergs/cm, and $K= 10^3~(5 \times 10^7$)~ergs/cc for the Fe (Sm-Co) layer.~\cite{FullertonPRB1998}  ($^\ast$Literature value of $K_{Fe}$ was used during the optimization.)}
\begin{center}
\begin{tabular}{l|ccccc}
\hline\hline
			                                          & Fe        &          & mixed    &         &Sm-Co  \\ \hline
Thickness  (nm)                                         & $7.4 \pm 0.3$       &          &  $4.2 \pm 0.3$               &          & $20.1 \pm 0.2$ \\
Roughness (nm)                                       &              & $4.3 \pm 0.3$   &                       &  $2.6 \pm 0.2$  &         \\
$\rho_{nuc}$    ($10^{-6}$~\AA$^{-2}$)& $7.9 \pm 0.2$    &          &$5.2 \pm 0.2$                  &           &  $1.64 \pm 0.04$     \\
$\rho_{mag}$  ($10^{-6}$~\AA$^{-2}$) &$4.9 \pm 0.1$      &           &$2.5 \pm 0.1$                 &            &$1.62 \pm 0.03$    \\
$M_s$    (emu/cc)        & $1700 \pm 50$      && $890 \pm 50$    & & $570 \pm 10$            \\ \hline
$M_{+0.39~T}$ (emu/cc)        & $1540 \pm 10$    && $730 \pm 30$  & & $420 \pm 10$          \\
$A$ (10$^{-6}$~ergs/cm)            &   $2.6 \pm 0.1 $               & &    $2.6 \pm 0.1 $         &  &  $1.2 \pm 0.1 $          \\
$K$ (10$^{6}$~ergs/cc)            & 0.001$^\ast$              & &  $3.0 \pm 0.2$       &  & $ 27 \pm 1 $      \\  \hline\hline
\end{tabular}%}
\end{center}
\end{table*}

The best-fit curves for the reflectivity data are overlaid on the data, and the SLD depth profiles  are shown in Fig.~\ref{Corefine}c. The SLD profiles show sharp transitions at the Cr/MgO and the Sm-Co/Cr interfaces with \emph{rms} roughnesses of $\sim$ 0.4 nm for both interfaces. However, the SLD profiles gradually change between the Fe and Sm-Co layers over a distance of $\sim 8$~nm. Such a large intermixed region is consistent with previous energy-dispersive-spectroscopy (EDS) results from samples fabricated under the same conditions.~\cite{ChoiPRB2007}  In the intermixed region, there appears a shoulder in the nuclear SLD profile (Fig.~\ref{Corefine}c), showing that it is chemically rich in Fe. However, this feature is not present in the magnetic SLD profile. This will be discussed below. The parameters of interest are listed in Table~\ref{tab:SLD},  which shows that both the Fe and the Sm-Co layers have their saturation magnetizations close to the literature values~\cite{FullertonPRB1998} despite the large intermixing at the interface.

\subsection{Depth Profiles of Exchange Stiffness and Uniaxial Anisotropy}
	\label{Sec:MM}	
\begin{figure}[b]
	\centering
		\includegraphics[width=0.48\textwidth]{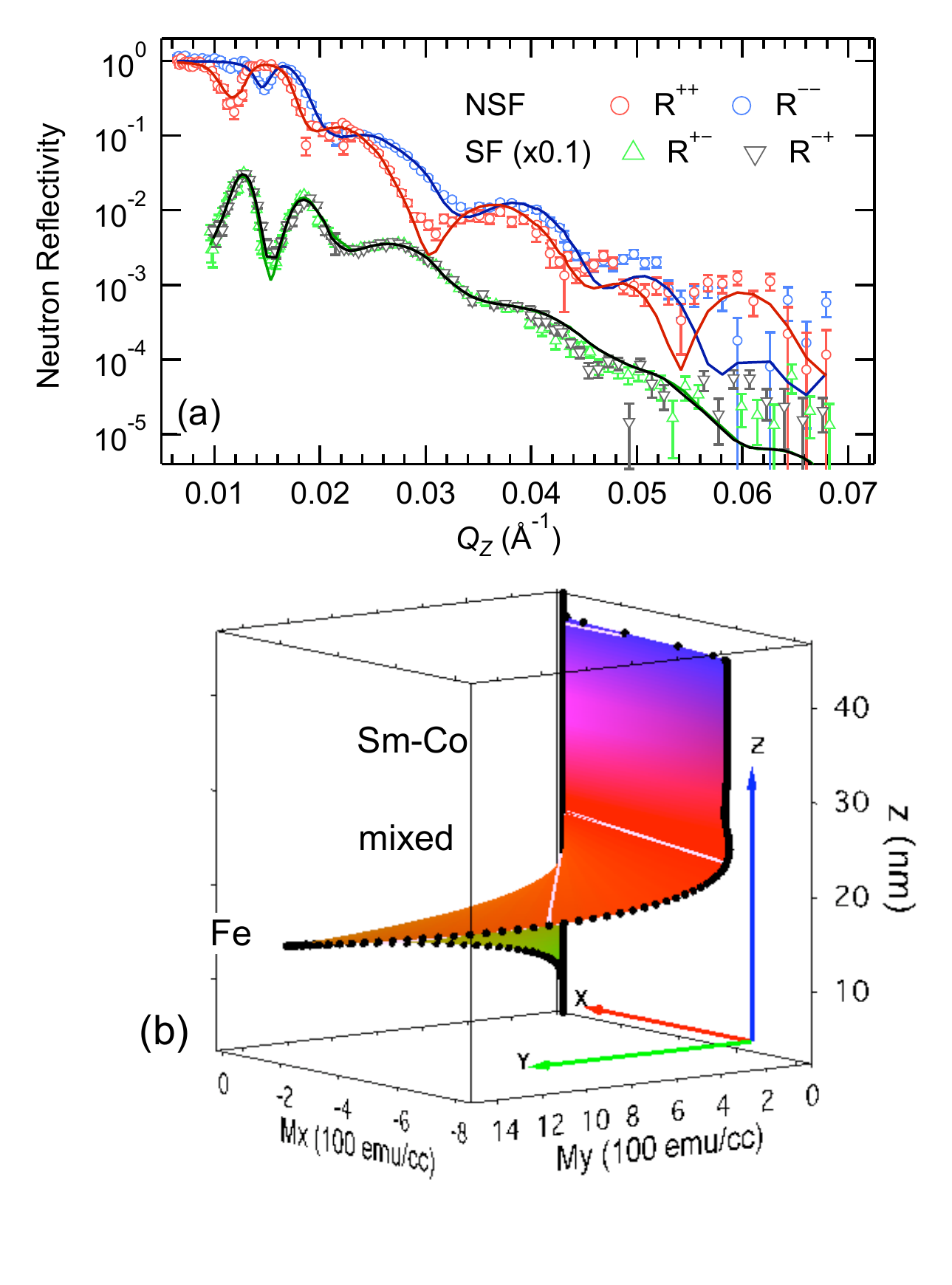}
	\caption{\label{MM3D} (Color online) (a)~PNR data taken at $+0.39$~T after saturation in $-1.11$~T.  $R^{+-}$ and $R^{-+}$ are offset by a factor of 0.1 for clarity. (b) The depth profiles of the magnetization vector $\vec M$ (obtained from micromagnetic simulations) that yields the best fit to the PNR data. The field direction is along $+ \hat x$.}
\end{figure}
In order to get insight into the intrinsic magnetic properties, besides $M_S$, in the intermixed region, we also studied the magnetization structure in a demagnetized state. The PNR data were collected at $+0.39$~T after the negative saturation in $-1.11$~T. The field is sufficiently high so that the specular SF reflection showed at off-specular positions on the detector due to the Zeeman effect, as shown in Fig.~\ref{fig:Zeeman}. Therefore, in the analysis the momentum transfer $Q_z$ was modified according to Eq.~\ref{eq:Zeeman}.~\cite{FelcherNature1995, FelcherPB1996}  It is worth noting that the off-specular scattering in the SF reflectivity is dominated by the Zeeman effect in our data and possible off-specular scattering due to in-plane magnetic domains is not distinguishable from the background, and is therefore not considered in the data.

The PNR data are shown in Fig.~\ref{MM3D}a. In contrast to the saturation case, $R^{--}$ is higher than $R^{++}$ for most $Q_z$ values above the critical edge, which indicates that the average $M_x$ is still along the negative applied field direction. Both the splitting between $R^{--}$ and $R^{++}$ and the separation between the dips below the critical edge are smaller because the average $M_x$ is much lower than $M_s$. At the same time, there is significant SF scattering of the same amplitude as the NSF scattering, indicating a large in-plane magnetization component perpendicular to the field direction ($M_y$).  This is caused by the magnetization spiral structure observed in exchange-coupled bilayers during demagnetization.~\cite{DonovanPRL2002} Figure~\ref{MM3D}b plots the depth profile of the magnetization vector $\vec M$ that yields the best fit, which is obtained from micromagnetic simulations. Our PNR studies do not reveal the chirality of $\vec M$ so that a positive $M_y$ is used for convenience. As expected, the magnetic moment of the Sm-Co layer is along the negative field direction and the magnetization vector rotates into the intermixed region and the Fe layer.

\begin{figure}[hb]
	\centering
		\includegraphics[width=0.45\textwidth]{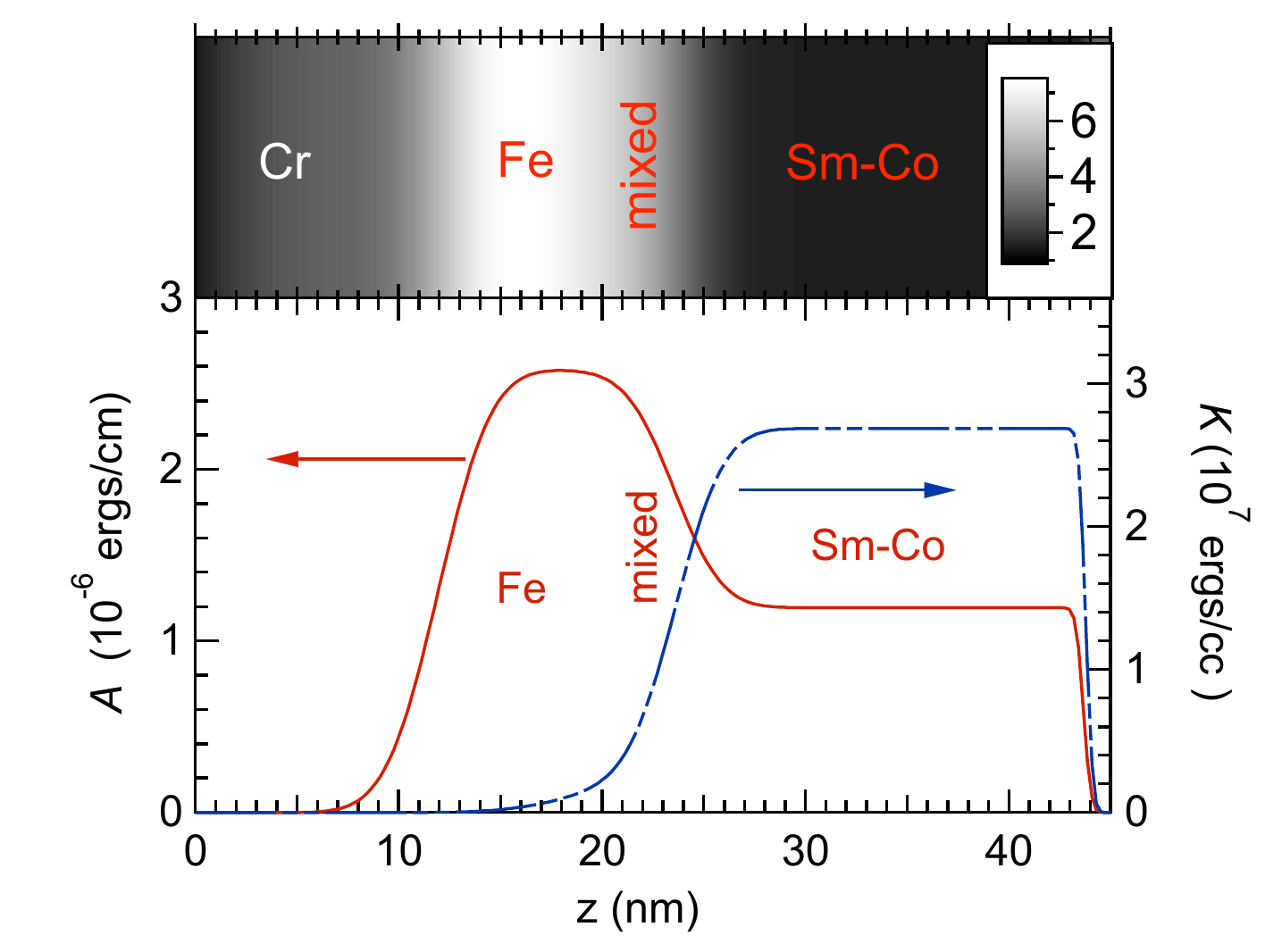}
	\caption{\label{AK} (Color online) Top: The depth profile of the nuclear SLD in gray scale in the unit of $10^{-6}$~\AA$^{-2}$. Bottom: The depth profile of the exchange stiffness $A$ (left, red solid line) and the uniaxial anisotropy $K$ (right, blue dashed line) which gives the best fit to the PNR data.}
\end{figure}

Rohlsberger \emph{et~al.} have shown that the depth profile of the magnetization vector in exchange coupled bilayers agrees with the micromagnetic simulation from the 1D spin-chain model.~\cite{RohlsbergerPRL2002} Fitzsimmons \emph{et~al.} further determined the micromagnetic parameters for individual layers in the exchange coupled DyFe$_22$/YFe$_2$ superlattice using combined studies of PNR and micromagnetic simulations.~\cite{FitzPRB2006} For the Fe/Sm-Co spring magnet studied here, the Fe and the Sm-Co layers are largely intermixed for $\sim 8$~nm at the interface and it is naturally to expect that the intrinsic magnetic properties, i.e. uniaxial anisotropy $K$ and exchange stiffness $A$, vary gradually due to the gradation of the chemical composition. Hence, we extended Fitzsimmons' approach and constructed depth profiles of the intrinsic magnetic properties. Given a profile of the micromagnetic parameters, the equilibrium magnetic structure can be simulated by energy minimization via the 1D spin-chain model,~\cite{FullertonPRB1998} from which the neutron's magnetic scattering potential can be directly calculated followed by the spin-dependent neutron reflectivities. Therefore, the intrinsic magnetic properties can be optimized to yield the best fit to the experimental PNR data.

These profiles of $K$ and $A$ are built up of layers identical to those of the chemical structure as determined from the fits to the XRR and the PNR data in saturations. The gradual changes between the layers are computed using the interface roughness in the same way as is done for the SLDs. When fitting, the chemical structure and the roughnesses are fixed to those already determined.   $M$ is allowed to be lower than $M_s$ to account for the effect from the minor soft phase mentioned above. The model optimization is not sensitive to the $K_{Fe}$ because the associated magnetic energy is negligible, and therefore the literature value of $K_{Fe}$ is used. The error bars of the parameters only reflect the statistical error and the accuracy of the values depends on the model. Therefore it is important here to have the additional layer between the Fe and the Sm-Co layers in the model because the exact mapping function from the chemical structure to the micromagnetic properties is unknown. The layer has its own free micromagnetic parameters so that $M$, $A$ and $K$ in the intermixed region are able to vary more independently from those in the Fe layer and the Sm-Co layer. We checked the robustness of the model by allowing different interface roughnesses for $M$, $A$, and $K$ and found that the optimized micromagnetic parameters have similar depth dependencies.

The depth profiles of $A$ and $K$ from the best fit are shown in Fig.~\ref{AK} and the values of $M$, $A$ and $K$ for layers of interest are listed in Table~\ref{tab:SLD}. Both the parameters $A$ and $K$ show monotonic  depth dependence as expected. $M_{Sm-Co}$ is $\sim 20\%$~lower than the saturation value, from which we estimated that $\sim 10\%$~of the Sm-Co was reversed due to the minor phase. Therefore, the 1D spin-chain model is approximately valid and yields fairly good fits, which were overlaid on the data in Fig.~\ref{MM3D}a. As also listed in Table~\ref{tab:SLD}, in the intermixed region, $A$ is close to that of the Fe layer, but $K$ is much lower than the average value between the Fe and the Sm-Co layers.

\section{Discussion and Summary}
\label{Discussion}

\begin{figure}[ht]
	\centering
		\includegraphics[width=0.45\textwidth]{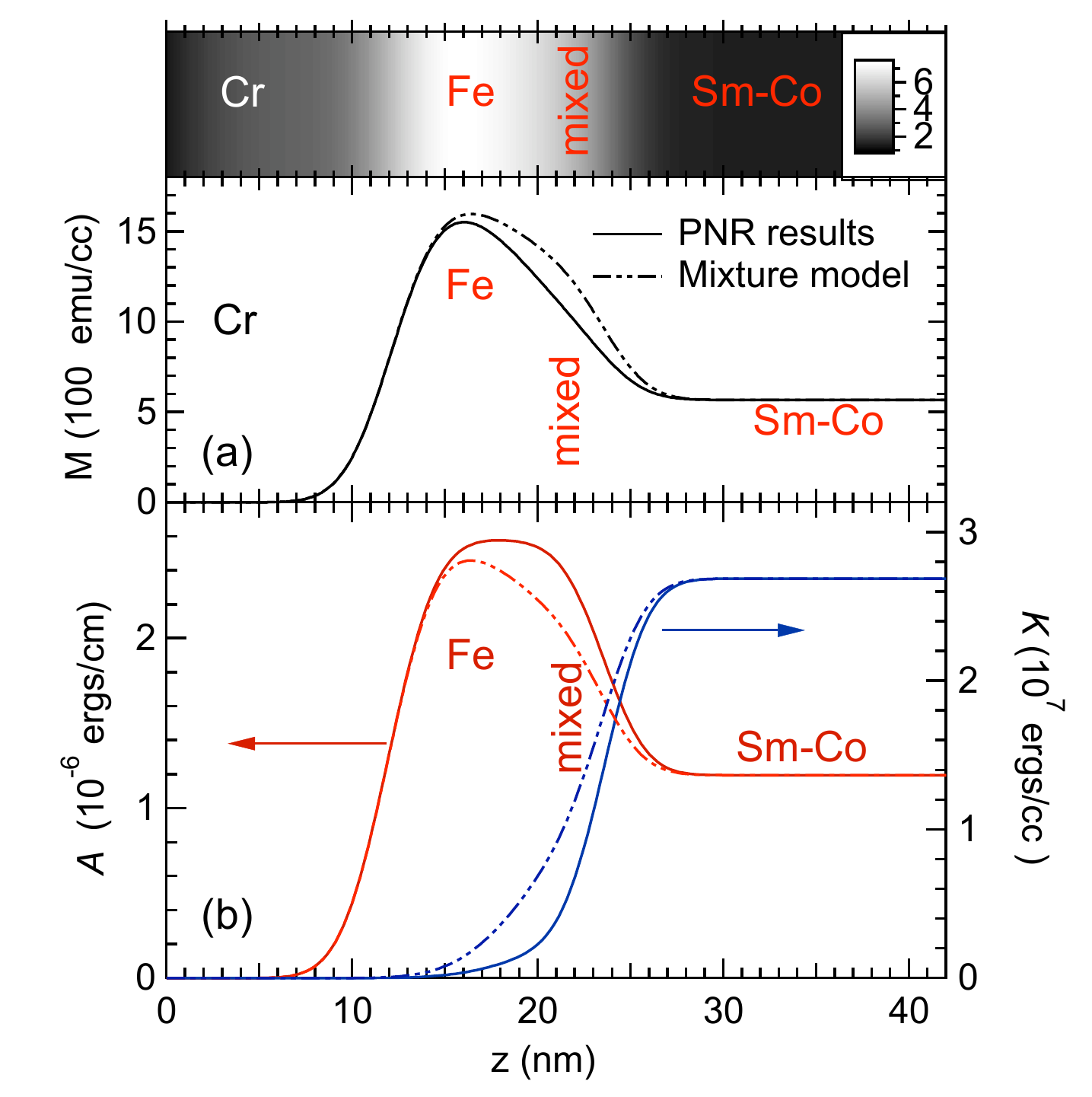}
	\caption{\label{Model} (Color online) Depth profiles of (a) the saturation magnetization, and (b) the exchange stiffness (left) and the uniaxial anisotropy (right) calculated by assuming that the intermixed region is a mixture of the Fe phase and the Sm-Co phase (dashed lines). The experimentally determined profiles are plotted as solid lines. Also shown in the top panel  is the depth profile of the nuclear SLD in gray scale. The unit is $10^{-6}$~\AA$^{-2}$. }
\end{figure}

By assuming that the intermixed region is composed of a mixture of the Fe and Sm-Co phases, the saturation magnetization can be computed from the relative volume of both phases via $M = f_s M_s + f_h M_h$, where $f_s$ ($M_s$) and $f_h$ ($M_h$) are the relative volume (the saturation magnetization) of the Fe phase and the Sm-Co phase, respectively. From the depth profile of the nuclear SLD, the relative volume of the two phases can be determined. Following this approach, the calculated  saturation magnetization profile is shown in Fig.~\ref{Model}a. Also shown is the saturation magnetization profile calculated from the magnetic SLD profile. Clearly, the mixture model predicted slightly larger saturation magnetization in the intermixed region.  This agrees with the suggestion that new compound(s) is (are) formed in the intermixed region,~\cite{JiangAPL2004} possibly due to interdiffusion between the Fe and Sm-Co layers. The estimation shows that the average magnetization of the whole sample is reduced by $\sim 4\%$ due to the intermixing of the two components. The exchange stiffness and the anisotropy of a two phase mixture can also be approximately estimated by the relative volume: $K = f_s K_s+f_h K_h$ and $A = f_s A_s+f_h A_h$.~\cite{SkomskiPRB1993} The results are shown in Fig.~\ref{Model}b.  In comparison to those determined from the PNR experiments, the model predicts higher anisotropy but lower exchange stiffness in the intermixed region. Therefore, we conclude that the optimized Fe/Sm-Co interface for a spring magnet reduces the average magnetization.

Interfaces with graded magnetic properties are desirable for superior performances in exchange coupling systems, such as spring magnets and magnetic media.~\cite{JiangAPL2004, VictoraIEEE2005} A straightforward approach to achieve magnetically graded interfaces is to make structurally graded interfaces.~\cite{ChoiAPL2007, GollJAP2008, KirbyPRB2010} The micromagnetic properties could be predicted from the composition. However, it is challenging if multiple phases coexist even when the micromagnetic properties of each individual phase are known, due to lacking of prior knowledge of the interphase coupling.  As shown above, the intrinsic magnetic properties in the intermixed Fe/Sm-Co interface deviates from the results predicted by the mixture model. Therefore, a complex correlation between the magnetic properties and the nominal composition may exist at the chemically graded interfaces.

In summary, we determined the intrinsic magnetic properties of a Fe/Sm-Co bilayer under the nearly optimized fabrication condition for spring magnets. We determined that the intermixed interface between the Fe and Sm-Co layers extends  $\sim 8$~nm, where the magnetic properties changes gradually, as expected. We compared the magnetic properties with the prediction of a mixture model and found, in the intermixed region, the saturation magnetization is slightly lower than that estimated from the model, but the exchange stiffness is higher. This observation indicates that the intermixed interface is efficient for magnetically coupling the Fe and Sm-Co layers but at the cost of the average magnetization. The intrinsic anisotropy is also lower than the value from the model. Overall, the intrinsic magnetic properties in the structurally intermixed region may not be predicted correctly by the mixture model of nominal compositions,  which is worth keeping in mind when designing the magnetically graded interfaces.

\begin{acknowledgments}
We thank Gian P. Felcher for helpful discussions. Research at Argonne was supported by the U.S. Department of Energy, Office of Basic Energy Sciences, Division of Materials Sciences and Engineering under Award No.DE-AC02-06CH11357. Research at Oak Ridge National Laboratory's Spallation Neutron Source was sponsored by the Scientific User Facilities Division, Office of Basic Energy Sciences, U. S. Department of Energy.
\end{acknowledgments}

\appendix
	\section{Zeeman effect and off-specular scattering}
	\label{App:Zeeman}

\begin{figure}[t]
	\centering
		\includegraphics[width=0.48\textwidth]{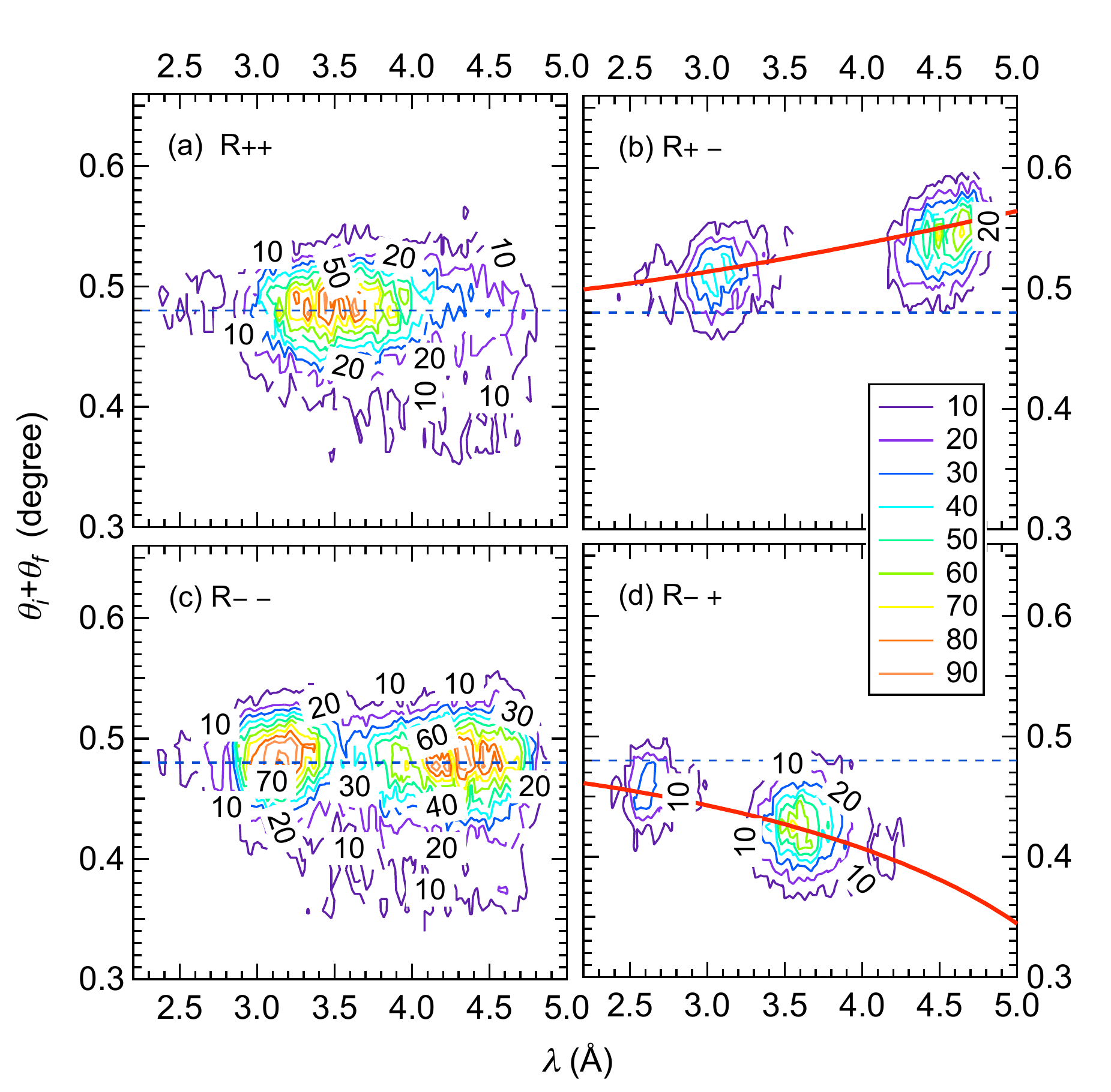}
	\caption{\label{fig:Zeeman} (Color online) The coutour map of the neutron intensities measured with the polarization analysis, which are presented as functions of the neutron wavelengths $\lambda$. The data were collected from a Fe/Sm-Co spring magnet sample at an incident angle $\theta_i = 0.24^o$ in an external field of +0.39~T after a negative saturation. The dashed lines indicate the specular reflection positions without considering the Zeeman effect, i.e. $\theta_f = \theta_i$. The solid lines in the SF channels indicate the off-specular scattering following the the Zeeman effect, i.e. $\sin^2 \theta_f =  \sin^2\theta_i \pm \frac{cB\lambda^2}{2\pi^2}$. }
\end{figure}

	The Zeeman effect in PNR was firstly clarified by Felcher~\emph{et. al}.~\cite{FelcherNature1995, FelcherPB1996} The effect shows up in the SF reflectivities when the difference of the Zeeman energy for spin-up and spin-down neutrons is not negligible in a sufficiently high magnetic field. Let $k_\perp$ and $k_{//}$ label the perpendicular and the parallel components of the wavevector with respect to the sample surface in the \emph{vacuum} region, \emph{i.e.}, before neutrons enter the magnetic-field region of interest. $k_\perp = 2 \pi sin\theta / \lambda \ll k_{//}$ at glancing angles. There is negligible energy exchange between the sample and the neutrons for elastic scattering, therefore the Zeeman energy change after SF scattering accompanies a change of  the kinetic energy, which is associated with $\vec k$. The energy change after spin-flip is twice the Zeeman energy. If the sample is magnetically homogenous in the film plane as seen by the neutrons, the energy change is totally associated with $k_\perp$. Since $k_\perp$'s are small quantities, the outgoing angles of the spin-flipped neutrons are different from the incoming angles. Therefore the SF reflections appear at off-specular positions, which have a characteristic field dependence.~\cite{FelcherNature1995, FelcherPB1996}  The Zeeman effect on the SF reflections becomes clear in a magnetic field on $\sim 0.1$~T or higher. Figure~\ref{fig:Zeeman} shows an example. It is clear that the locus of the SF reflections deviates away from the specular position, but are described by the prediction after considering the Zeeman effect. Beside the reflection due to the Zeeman effect, there is no noticeable off-specular scattering in the SF channels; therefore, the scattering from the in-plane magnetic domains are not considered during the data analysis. Consequently, the momentum transfer $Q_z$ are no longer equal to $2k_\perp$ for spin-flipped neutrons, but follows
\begin{equation}
 \label{eq:Zeeman}
\begin{array}{rcl}
Q_z^{+-} & =  &k _\perp + \sqrt{k_\perp^2 + 2  C B} \\
Q_z^{-+} & =  &k_\perp + \sqrt{k_\perp^2 - 2 C  B},
\end{array}
\end{equation}
where $C=| 2m_n \mu_n / \hbar ^2| = 2.906 \times 10^{-5}$~ \AA$^{-2}$/T and $B = \mu_0 H$ is the applied magnetic field.  $R^{-+}$ is forbidden when $k_\perp < \sqrt{2CB}$ because there is no enough kinetic energy to compensate the Zeeman energy change.~\cite{vandeKruijsPB2000} At the same time, the minimum $Q_z^{+-}$ is also $\sqrt{2CB}$ for finite reflectivity. Therefore, there is a cutoff $Q_z$ for non-zero SF scattering, $Q_{SFcutoff} =  \sqrt{2CB}$.

	Actually, the kinetic energy also changes when neutrons enter and leave the magnetic-field region of interest, but the change of $\vec k$ is essentially in $k_{//}$ in these cases,~\cite{Chatterji2006} which changes the time-of-flight. However, the relative change of $k_{//}$ is so small in laboratory fields, and TOF only changes $\sim 10$~ns, while the width of the neutron pulse is $\sim 0.2$~ms. Therefore, it does not result in any observable affect.

	SF reflection is typically weak when the Zeeman effect matters, therefore the Zeeman effect is usually not considered during data reduction and analysis.~\cite{Reflpak, Fitz2005} However, we observed significant SF scattering at $+0.39$~T, which is sufficiently high so that the Zeeman effect needs to be considered. We adopted the generalized algorithm of the GEPORE~\cite{Chatterji2006} for the PNR simulations. $\vec B$ rather than $\vec M$ is used to calculate the magnetic scattering potential since $H$ is now comparable to $M$ and the spin eigenstates are aligned along $\vec B$ rather than $\vec M$. Considering the problem along the normal direction of the sample surface (let it be $\hat z$), it is reduced to a pair of coupled 1D differential wave equations,~\cite{Chatterji2006}
\begin{equation}
 \label{eq:1Dcoupled}
 \begin{array}{rcl}
\left[ -\frac{\hbar^2}{2m}k_0^2 + V_{++}(z) - E \right]  \Psi_+(z) + V_{+-}\Psi_-(z)  & =  & 0  \\
\left[ -\frac{\hbar^2}{2m}k_0^2 + V_{--}(z)  - E \right]   \Psi_-(z) + V_{-+}\Psi_+(z)  & =  & 0.
\end{array}
\end{equation}
 $V$ is the potential operator. $\Psi_{\pm}(z)$'s are the wave functions for spin-up and spin down states. $k_0$ is the wavevector in vacuum, which can be either a real number or a pure imaginary number. States with ${k_0}^2 \pm CB < 0$ correspond to evanescent waves outside the sample. Due to the Zeeman energy, the coupled eigenstates of $\Psi_{+}$ and $\Psi_{-}$ have different wavevectors outside the sample, namely $\sqrt{{k_0}^2 - CB}$ and $\sqrt{{k_0}^2 + CB}$, which has been considered to set up the boundary conditions at the sample surface in the modified algorithm to take account the Zeeman effect.

\bibliography{FeSmCo}

\begin{thebibliography}{10}
\bibitem{Model}%
  \BibitemOpen
  \bibinfo {note} {It was found that the PNR results suggest a larger surface
  roughness and a thicker top Cr layer than XRR. The thickness difference is
  about 0.9 nm and is most likely due to the increase of the oxide layer after
  $\sim 2$~years.}%
  \bibAnnoteFile{Stop}{Model}%
\end{thebibliography}

\end{document}